\documentclass[preprint,5p]{elsarticle}

\usepackage{amsmath}
\usepackage{amsthm}
\usepackage{amssymb}
\usepackage{bm} 

\usepackage{hyperref}
\usepackage{lineno}

\usepackage{color}

\allowdisplaybreaks

\journal{Physics Letter B}

\begin{document}

\begin{frontmatter}

\title{Exact-exchange relativistic density functional theory in three-dimensional coordinate space}

\author[RCNP,CENS,PKU]{Qiang Zhao}
\author[JCHP,PKU]{Zhengxue Ren}
\author[PKU]{Pengwei Zhao\corref{cor1}}
\ead{pwzhao@pku.edu.cn}
\author[RCNP,RNC,CCS]{Kenichi Yoshida}

\cortext[cor1]{Corresponding author}

\address[RCNP]{Research Center for Nuclear Physics, Osaka University, Ibaraki, Osaka 567-0047, Japan}
\address[CENS]{Center for Exotic Nuclear Studies, Institute for Basic Science, Daejeon 34126, South Korea}
\address[PKU]{State Key Laboratory of Nuclear Physics and Technology, School of Physics, Peking University, Beijing 100871, China}
\address[JCHP]{Institute for Advanced Simulation, Forschungszentrum J\"{u}lich, D-52425 J\"{u}lich, Germany}
\address[RNC]{RIKEN Nishina Center for Accelerator-Based Science, Wako, Saitama 351-0198, Japan}
\address[CCS]{Center for Computational Sciences, University of Tsukuba, Tsukuba, Ibaraki 305-8577, Japan}

\begin{abstract}
The exact-exchange relativistic density functional theory (Ex-RDFT) of atomic nuclei has been solved in three-dimensional lattice space for the first time. 
The exchange energy is treated within the framework of the orbital-dependent relativistic Kohn-Sham density functional theory, wherein the local Lorentz scalar and vector potentials are derived using the relativistic optimized effective potential method.
The solutions of binding energies, charge radii, and density distributions are benchmarked against the traditional relativistic Hartree-Fock approach for spherical and axially deformed nuclei. 
Furthermore, the triaxial neutron-rich $^{104-120}\text{Ru}$ isotopes are investigated with the exchange correlations, which is beyond the current capacity of the traditional relativistic Hartree-Fock approach. 
The results notably indicate the $\gamma$-softness of these neutron-rich nuclei, which is consistent with experimental observations. 
This novel approach establishes a foundation for the study of nuclei without imposing any symmetry restrictions employing relativistic density functional with exchange correlations.
\end{abstract}


\end{frontmatter}



The worldwide development of rare-isotope beam facilities brings us various intriguing phenomena about nuclear shapes including the shape phase transitions \cite{cejnar2010_RMP82_2155} and the shape coexistences \cite{heyde2011_RMP83_1467, garrett2021_PiPaNP_103931}.
All these phenomena are closely tied to the evolution of the single-nucleon shell structure, in which the spin-orbit and tensor interactions \cite{otsuka2020_RMP92_015002} play an important role.
Continuum effects can also be crucial for weakly bound nuclei near the dripline, which should be properly considered \cite{Dobaczewski1994_PRL72.981, Dobaczewski1996_PRC53.2809,vretenar2005_PR409_101,meng2006_PiPaNP57_470}.
Therefore, a unified microscopic theory is needed for a comprehensive understanding of the various novel phenomena associated with nuclear shapes.  

Density functional theory (DFT) is the sole microscopic theory that can provide a self-consistent description for nuclei across the entire nuclear chart \cite{bender2003_RMP75_121, Meng2016_book}.
In particular, the Kohn-Sham DFT offers a practical way to construct the energy density functional by introducing an auxiliary noninteracting system where particles move in a local potential, generating the same ground-state density as in the interacting system \cite{kohn1965_PR140_A1133}.
This reduces the complicated quantum many-body problem to a nonlinear one-body problem.

The relativistic density functional theory (RDFT) \cite{Serot1986Adv.Nucl.Phys.1,ring1996_PiPaNP37_193,vretenar2005_PR409_101,meng2006_PiPaNP57_470} has attracted a lot of attention as it leverages the Lorentz symmetry \cite{ring2012_PST150_014035}.
This allows the RDFT to naturally take into account the spin degrees of freedom, resulting in large spin-orbit splittings, which are evident in the nonrelativistic reduction of the RDFT via a similarity renormalization method~\cite{Ren2020SelfSRG}.  
It also guarantees the consistent treatment of the time-odd fields without requiring additional parameters.
The success of the RDFT has been demonstrated through its extensive applications in describing various static and dynamic properties of finite nuclei \cite{ring1996_PiPaNP37_193,vretenar2005_PR409_101,meng2006_PiPaNP57_470,niksic2011_PiPaNP66_519,ren2020_PLB801_135194,ren2022_PRC105_L011301,ren2022_PRL128_172501}.
Recent works have also revealed the importance of Lorentz symmetry in describing nuclear force \cite{Lu2022_PRL128.142002}, nuclear binding \cite{Yang2022_PLB835.137587}, and nuclear neutrinoless double beta decay \cite{Yang2024_PLB855.138782} within the effective field theory. 

It is important to note that the majority of RDFTs comprise solely of the kinetic and Hartree energies, while the exchange energies are usually neglected for simplicity.
Although it is assumed that some of the exchange correlations could be captured by adjusting the coupling constants in the density functionals, some shortcomings have been identified when these theories are employed to investigate the evolution of nuclear shell structures \cite{Long2007Phys.Rev.C34314, long2008_EEL82_12001,wang2013_PRC87_047301, Shen2018Phys.Lett.B344} and the charge-exchange spin-flip excitations \cite{liang2008_PRL101_122502,liang2012_PRC85_064302}.
In particular, the $\pi$-exchange, which is a primary source of the tensor force, is absent due to its negative-parity nature.

One direct method for incorporating the exchange energies is the relativistic Hartree-Fock (RHF) approach.
With the $\pi$-exchange and $\rho$-tensor couplings, which can be naturally introduced in the exchange energies, the RHF approach has successfully described the new shell structures \cite{li2016_PLB753_97, liu2020_PLB806_135524} and the proton bubble structure \cite{Li2019_PLB788.192}.
Furthermore, it has been demonstrated that the RHF approach can self-consistently reproduce the excitation energy of spin-isospin resonances without any additional adjustments \cite{liang2008_PRL101_122502, liang2012_PRC85_064302}.
However, the complicated nonlocal potentials involved in the RHF theory lead to much higher computational costs, in particular for deformed nuclei~\cite{geng2020Phys.Rev.C101.064302,geng2022Phys.Rev.C105.034329}.
To date, the RHF theory is limited to solving nuclei with spherical and axial symmetries. 
The RHF calculations based on the Dirac Woods-Saxon basis expansion for axially deformed nuclei require a large amount of computational resources. 
Consequently, it is advantageous to maintain the Kohn-Sham framework that contains only local potentials.

By employing contact interactions, the exchange terms can be represented as Hartree terms via the Fierz transformation~\cite{liang2012_PRC86_021302R,zhao2022_PRC106_034315} and, thus, the RHF framework is much simplified, while the exchange of light $\pi$ mesons cannot be considered since it is associated with the long-range dynamics.
Recently, an exact-exchange relativistic density function theory (Ex-RDFT) has been proposed within the Kohn-Sham scheme~\cite{zhao2023_PLB841_137913}. This approach incorporates relativistic exchange energies by constructing an orbital-dependent relativistic density functional.
The fully local relativistic Kohn-Sham (RKS) potentials are determined using the relativistic optimized effective potential (ROEP) method~\cite{shadwick1989_CPC54_95,engel1995_PRA52_2750,engel1998_PRA58_964,kodderitzsch2008_PRB77_045101a}. 
The validity and efficiency of the Ex-RDFT have been demonstrated for spherical nuclei through comparison with RHF calculations.

Since most nuclei are deformed, an extension of the Ex-RDFT to the deformed case is essential. 
As only local potentials are involved in the Ex-RDFT, it is relatively straightforward to employ the modern techniques for solving the Dirac equation in three-dimensional (3D) coordinate space \cite{ren2017_PRC95_024313,li2020_PRC102_044307}. 
This allows for the description of arbitrary deformation. 
To solve the Dirac equation in 3D coordinate space faces two main problems: the variational collapse \cite{zhang2010_IJMPE19_55} and the Fermion doubling problems \cite{tanimura2015_PoTaEP2015_073D01}. 
The variational collapse problem can be addressed by the inverse Hamiltonian method \cite{hagino2010_PRC82_057301} or the preconditioned conjugate gradient method with a filtering function \cite{li2020_PRC102_044307}.
The Fermion doubling problem can be resolved by adopting the Fourier spectral method \cite{ren2017_PRC95_024313}.
The RDFT in 3D coordinate space with only Hartree energy has been used to study the exotic nuclear shapes, such as linear chain structures~\cite{ren2019_SCPMA62_112062,Zhang2022PRC_LCS}, toroidal states~\cite{ren2020_NPA996_121696}, tetrahedral shapes~\cite{xu2024_PRC109_014311,Xu2024PLB_tetrahedral}, and high-order octupole and hexacontetrapole deformations~\cite{Xu2024PRL_emergence}.

In this work, to incorporate the exact exchange energy for nuclei with arbitrary deformation, the Ex-RDFT will be solved in 3D coordinate space using the inverse Hamiltonian and spectral methods. 
The new framework will be validated under both spherical and axial symmetries by benchmarking against the RHF results. 
Following validation, it will be applied to investigate the shape evolution in neutron-rich Ru isotopes, incorporating the triaxial degree of freedom.
This provides the first study of triaxial nuclei with the relativistic exchange correlation energies. 

%
The energy density functional is written as the RHF energy, which is derived from an effective Lagrangian. In this Lagrangian, the nucleons interact with the $\sigma$, $\omega$, and $\rho$ mesons as well as the photons \cite{bouyssy1987_PRC36_380,long2006_PLB640_150}.
The energy density functional is composed of three parts
\begin{align}\label{equ:edf}
 E = T + E_{\rm H} + E_{\rm x},
\end{align}
where the kinetic energy $T$, the Hartree energy $E_{\rm H}$, and the exchange energy $E_{\rm x}$ are defined in terms of single-particle states $\{\varphi_a\}$,
\begin{align}
T =& \sum_a v_a^2 \int d\bm r ~ 
    \bar\varphi_a[-i\bm\gamma\cdot\bm\nabla+M]\varphi_a, \\
E_{\rm H} =& \frac{1}{2}\sum_{\phi;ab} v_a^2 v_b^2 \iint d\bm r d\bm r'
    \left[\bar\varphi_a\Gamma_\phi\varphi_a\right]_{\bm r} \nonumber \\
    & \qquad\qquad\qquad\qquad\quad \times D_\phi(\bm r,\bm r')
    \left[\bar\varphi_b\Gamma_\phi\varphi_b\right]_{\bm r'}, \\
E_{\rm x} =& -\frac{1}{2}\sum_{\phi;ab} v_a^2 v_b^2 \iint d\bm r d\bm r'
    \left[\bar\varphi_a\Gamma_\phi\varphi_b\right]_{\bm r} \nonumber \\
    & \qquad\qquad\qquad\qquad\quad \times D_\phi(\bm r,\bm r')
    \left[\bar\varphi_b\Gamma_\phi\varphi_a\right]_{\bm r'}.
\end{align}
Here, $M$ is the nucleon mass, $v_a^2$ is the occupation probability, $\Gamma_\phi$ and $D_\phi$ are the interaction vertex and the propagator, respectively, with $\phi$ denoting $\sigma$, $\omega$, $\rho$ mesons and the photon $A$.
Note that $T$ and $E_{\rm x}$ are orbital-dependent functionals, while $E_{\rm H}$ can be expressed as a functional of the scalar density $\rho_{s}$ and the vector currents $j^\mu$:
\begin{align}
  \rho_{s,\tau}(\bm r)
    =&\sum_{a\in \tau} v_a^2 
    \bar\varphi_{a}(\bm r)
    \varphi_{a}(\bm r), \\
  j^\mu_\tau(\bm r)
    =&\sum_{a\in \tau} v_a^2 
      \bar\varphi_{a}(\bm r)
      \gamma^\mu
      \varphi_{a}(\bm r).
\end{align}
with $\tau$ representing neutrons or protons.

In contrast to the RHF theory, the single-particle states in Ex-RDFT, following the Kohn-Sham procedures, are defined by solving the RKS equation, which is essentially a Dirac equation,
\begin{equation}\label{equ:RKS_equation}
  \hat{h}\varphi_{a}(\bm r)
=\varepsilon_{a}\varphi_{a}(\bm r),
\end{equation}
where $\hat{h}$ is the single-particle Hamiltonian, 
\begin{equation}
  \hat{h} =
  -i\bm{\alpha\cdot\nabla}
      +\beta\left[
      M+S_\tau(\bm r)+\gamma_\mu V^\mu_\tau(\bm r)
  \right].
\end{equation}
Note that $S_{\tau}$ and $V_{\tau}^\mu$ are the local RKS potentials derived from the variation of the energy functional, which can be further divided into Hartree and exchange components,
\begin{align}
    S_\tau (\bm r)&
  = S_{\rm H,\tau} (\bm r) + S_{\rm x,\tau}(\bm r)
  =\frac{\delta E_{\rm H}}{\delta \rho_{s,\tau}}
  +\frac{\delta E_{\rm x}}{\delta \rho_{s,\tau}}, \\
  V_\tau^\mu (\bm r)&
  = V^\mu_{\rm H,\tau}(\bm r) + V^\mu_{\rm x,\tau}(\bm r)
  =\frac{\delta E_{\rm H}}{\delta (j_\tau)_\mu}
  +\frac{\delta E_{\rm x}}{\delta (j_\tau)_\mu}.
\end{align}
The Hartree potentials $S_{\rm H,\tau} (\bm r)$ and $V^\mu_{\rm H,\tau}(\bm r)$ are straightforward to obtain due to the explicit dependence of the Hartree energies on densities and currents.
However, the exchange potentials $S_{\rm x,\tau} (\bm r)$ and $V^\mu_{\rm x,\tau}(\bm r)$ are more complicated due to the orbital-dependent nature of the exchange energy.   
The general procedure to get the exchange RKS potentials is solving the ROEP equations, derived via the chain rule of functional differentiation \cite{zhao2023_PLB841_137913},
\begin{align}
  &S_{\rm x,\tau} \rho_{s,\tau} + V^\nu_{\rm x,\tau} (j_{\tau})_\nu  
  =\frac{1}{2}\sum_{a\in\tau} v_a^2
    \bigg\{
      \bar\varphi_a  (W_a -\zeta_{a} \gamma^0)  \varphi_{a} + {\rm c.c.}
    \bigg\}, \label{equ:ROEP1} \\
  &S_{\rm x,\tau}j^\mu_{\tau} + V^\mu_{\rm x,\tau} \rho_{s,\tau} 
  = \frac{1}{2}\sum_{a\in\tau} v_a^2
    \bigg\{
      \bar\varphi_a (W_a -\zeta_{a} \gamma^0)  \gamma^\mu\varphi_{a} + {\rm c.c.}
    \bigg\},
    \label{equ:ROEP2} 
\end{align} 
with the orbital-specific potential $W_a=\frac{1}{\bar\varphi_a}\frac{\delta E_{\rm x}}{\delta\varphi_a}$ and the energy shift $\zeta_a$. Similar to Ref. \cite{zhao2023_PLB841_137913}, the orbital-shift terms are neglected for simplicity by adopting the relativistic Krieger-Li-Iafrate (RKLI) approximation \cite{kreibich1998_PRA57_138, krieger1990_PLA146_256}, which has been shown to be a good approximation via some typical spherical nuclei.

To describe nuclei without any symmetry restrictions, we solve the RKS equation (\ref{equ:RKS_equation}) in 3D coordinate space.
The single-particle wave function $\varphi_a$ is solved iteratively by the inverse Hamiltonian method \cite{hagino2010_PRC82_057301,ren2017_PRC95_024313},
\begin{equation}
   \varphi_a^{(n+1)} = \mathcal{O}\left\{
    \left(1+\frac{\Delta t}{\hat{h}-Q_a}\right)\varphi_a^{(n)}
    \right\},
\end{equation}
where $\mathcal{O}$ denotes the orthonormalizaiton of the wave functions, $n$ is the iterative number, $Q_a$ is the shift parameter, and $\Delta t$ is the iterative step size.
Since the Ex-RDFT contains only local potentials, the application of the inverse Hamiltonian method in our framework is quite straightforward as compared to the RHF equation that involves nonlocal exchange potentials.
To avoid the fermion doubling problem when using the finite differences, the spectral method is applied for the involved spatial derivatives~\cite{ren2017_PRC95_024313}.
In particular, the spectral method facilitates the calculation of the exchange energy through the convolution theorem:
\begin{align}
  E_{\rm x} =& -\frac{1}{2}\sum_{\phi;ab} v_a^2 v_b^2 \int d\bm r
    \left[\bar\varphi_a\Gamma_\phi\varphi_b\right]_{\bm r} \nonumber \\
    &\qquad\qquad\quad \times
      \mathcal{F}^{-1}\left[
      \mathcal{F}\left(D_\phi\right)
      \mathcal{F}\left(\left[\bar\varphi_b\Gamma_\phi\varphi_a\right]_{\bm r'}\right)
    \right],
\end{align}
where $\mathcal{F}$ and $\mathcal{F}^{-1}$ represent the Fourier and inverse Fourier transformation, respectively.

In this work, the optimized effective interaction PKO2 \cite{long2008_EEL82_12001} is taken as the two-body interaction in the energy density functional.
The microscopic center-of-mass correction \cite{bender2000_EA7_467,long2004_PRC69_034319,zhao2009_CPL26_112102} is taken into account for the total energy. 
The BCS method is employed to consider the pairing correlations, where a delta pairing force is used with smooth energy-dependent cut-off weights~\cite{bender2000_EPJA8_59a},
\begin{align}
  f_a = \frac{1}{1+\exp[(\varepsilon_a-\lambda_\tau - \Delta\epsilon_\tau)/d_\tau]},
\end{align}
in which $\lambda_\tau$ is the Fermi energy, and the cut-off parameters are taken $\Delta\epsilon_\tau = 5 \text{~MeV}$ and $d_\tau = \Delta\epsilon_\tau/10$.

Our method is benchmarked against the spherical and axially-deformed RHF models, which are solved using the spherical Dirac Woods-Saxon basis.
Since the finite-range Gogny force D1S \cite{berger1984_NPA428_23} is adopted for the $pp$ channel in the RHF+BCS calculations, the strengths of the delta pairing forces here are adjusted to reproduce the average pairing gaps given by the RHF+BCS calculations for spherical shapes.
The coordinate space is discretized by 26 grid points with step sizes 1.0 fm along the $x$, $y$, and $z$ directions.
The size of the space has been verified to be sufficient for the numerical convergence.
The iterative step size $\Delta t$ is set to 100 MeV.
The convergence criteria for the iteration require that the energy uncertainties, defined as $\langle\hat{h}^2\rangle-\langle\hat{h}\rangle^2$, of all occupied states are less than $10^{-9}~\rm{MeV}^2$ and the maximum change in densities between two iterations is less than $10^{-7}$ fm$^{-3}$.


\begin{table}[!htbp]
\centering
\caption{The ground-state energies and charge radii of the spherical nuclei $^{16}$O, $^{48}$Ca, $^{132}$Sn, $^{208}$Pb and the deformed nucleus $^{20}$Ne calculated by the Ex-RDFT with the effective interaction PKO2 \cite{long2008_EEL82_12001}. The RHF results are shown for comparison.}
\label{tab:berc}
\begin{tabular}{cccccc}
\hline\hline
& \multicolumn{2}{c}{Ex-RDFT} & &  \multicolumn{2}{c}{RHF} \\ \cline{2-3} \cline{5-6}
      &  $E_B$ (MeV)  & $R_c$ (fm)  & &  $E_B$ (MeV)  & $R_c$ (fm) \\ \hline
$^{16 }$O  &  -126.63 & 2.729 & &  -126.76 & 2.715  \\ 
$^{48 }$Ca &  -415.02 & 3.493 & &  -415.15 & 3.474  \\ 
$^{132}$Sn & -1102.37 & 4.751 & & -1102.60 & 4.720  \\ 
$^{208}$Pb & -1635.68 & 5.555 & & -1636.01 & 5.519  \\ \hline
$^{20 }$Ne &  -155.38 & 2.977 & &  -155.56 & 2.957  \\ 
\hline\hline
\end{tabular}  
\end{table}

\begin{figure}[htbp]
  \centering
  \includegraphics[width=0.44\textwidth]{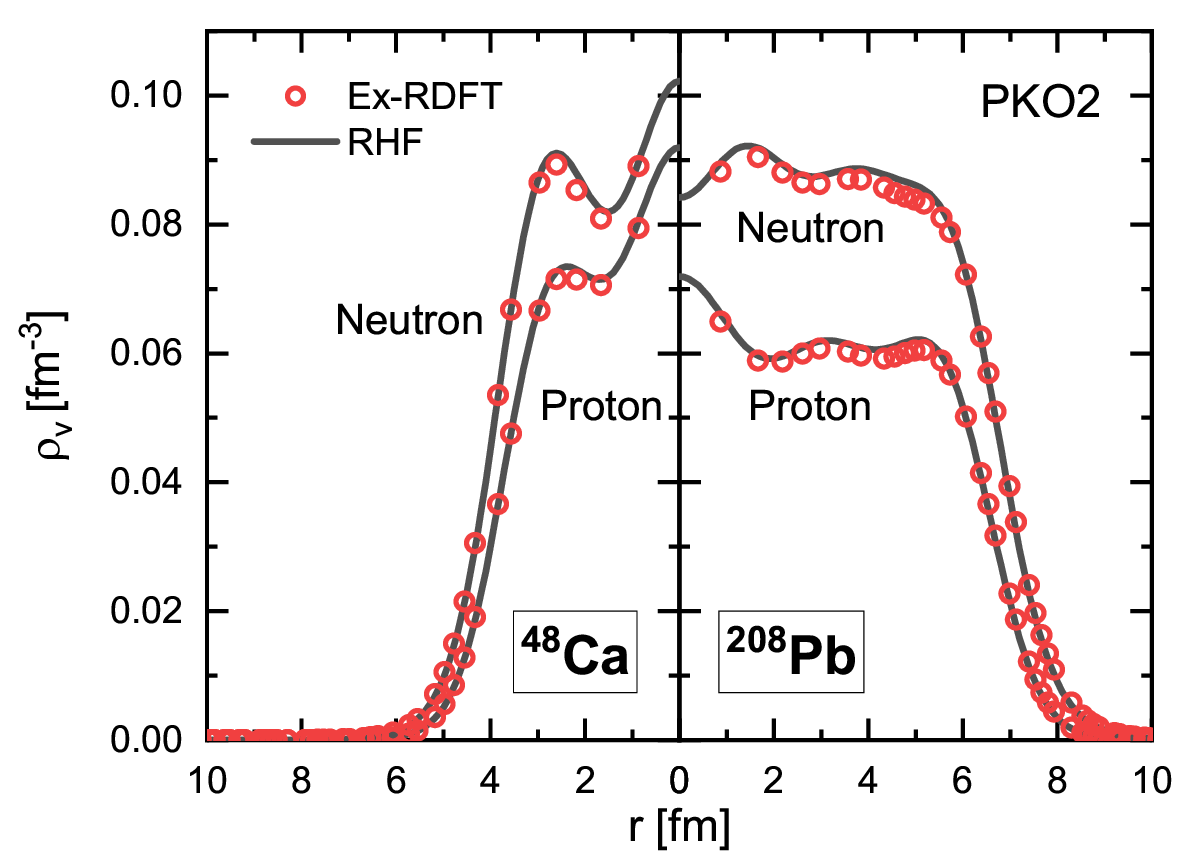}
  \caption{(Color online).
  Neutron and proton density distributions of $^{48}$Ca and $^{208}$Pb obtained by the Ex-RDFT in 3D coordinate space with the effective interaction PKO2 in comparison with the RHF results.
  }
  \label{fig:density}
\end{figure}

In Table \ref{tab:berc}, the ground-state energies and charge radii for the spherical nuclei $^{16}$O, $^{48}$Ca, $^{132}$Sn, and $^{208}$Pb given by the Ex-RDFT in 3D coordinate space are compared with the corresponding RHF results.
The Ex-RDFT results are in good agreement with the RHF results, with the differences of less than 0.33 MeV in the binding energy and 0.036 fm in the charge radius for the heavy nucleus $^{208}\rm{Pb}$.
The present Ex-RDFT results are also consistent with the previous Ex-RDFT results obtained under spherical symmetry in Ref. \cite{zhao2023_PLB841_137913}.

In accordance with the Kohn-Sham DFT concept, the ground-state density distributions yielded by the Ex-RDFT should be identical to those obtained by RHF.
The calculated density distributions of $^{48}$Ca and $^{208}$Pb are depicted in Fig. \ref{fig:density}.
The Ex-RDFT results demonstrate a notable concordance with the RHF results. 
 
\begin{figure}[htbp]
  \centering
  \includegraphics[width=0.40\textwidth]{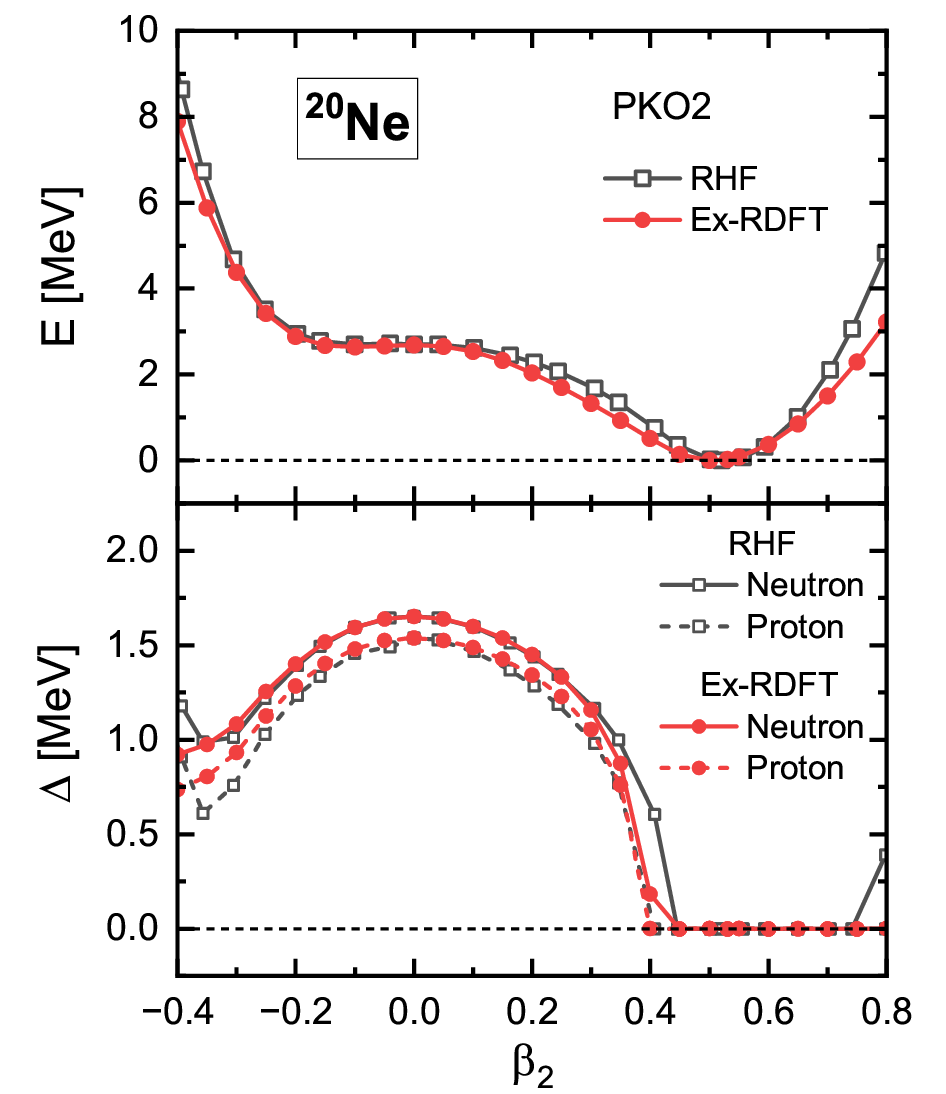}
  \caption{(Color online).
  Potential energy curve against the quadrupole deformation $\beta_2$ (upper panel) and average pairing gap (lower panel) for $^{20}$Ne obtained by Ex-RDFT with the effective interaction PKO2 in comparison to RHF results.
  }
  \label{fig:Ne20}
\end{figure}

By taking the axially deformed nucleus $^{20}$Ne as an example, we benchmark the Ex-RDFT calculations for deformed nuclei against the corresponding RHF calculations \cite{geng2020Phys.Rev.C101.064302}.  
The upper panel of Fig. \ref{fig:Ne20} depicts the potential energy curves along the quadrupole deformation parameter $\beta_2$ for $^{20}$Ne. 
The energies here are normalized with respect to the ground-state energy.

The results obtained with the Ex-RDFT are in general agreement with those obtained from the RHF method.
The ground state of $^{20}$Ne has a large prolate deformation at $\beta=0.531$, which is close to the RHF value of 0.520.
The discrepancy in the ground-state total energy between the Ex-RDFT and RHF calculations is 0.18 MeV, and the difference in charge radius is 0.020 fm as seen in Table \ref{tab:berc}.
The pairing force has been adjusted to reproduce the RHF average pairing gap at $\beta_2=0.0$.
As depicted in the lower panel of Fig. \ref{fig:Ne20}, the average pairing gaps obtained by the Ex-RDFT calculations are generally consistent with those from the RHF calculations.
The Ex-RDFT also reproduces the second minimum, characterized by a soft oblate deformation around $\beta_2=-0.1$, approximately 2.5 MeV above the ground state.
This demonstrates that the Ex-RDFT aligns well with the RHF results for deformed nuclei.

\begin{figure*}[!htbp]
  \centering
  \includegraphics[width=0.85\textwidth]{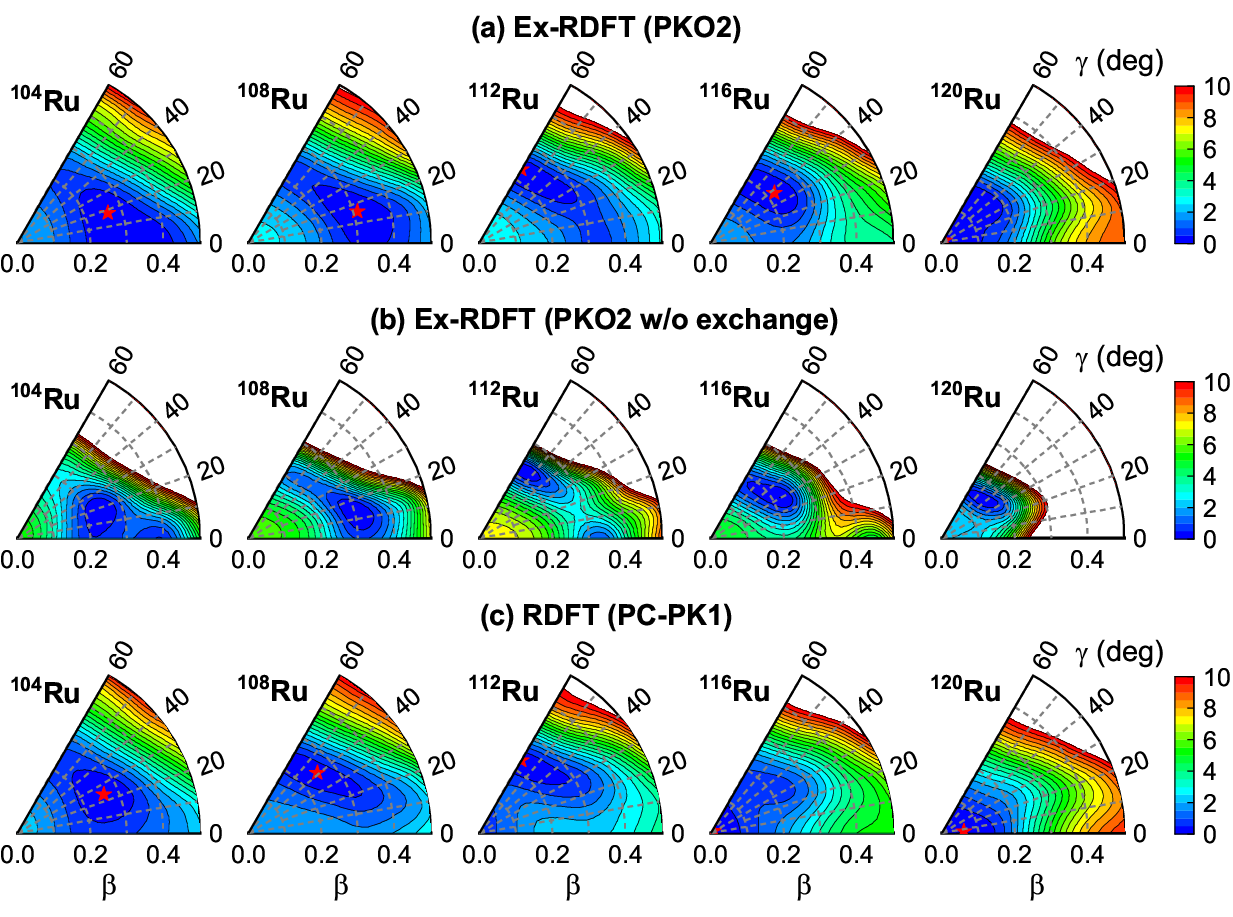}
  \caption{(Color online).
  Potential energy surfaces (PESs) of Ru isotopes in the $\beta$-$\gamma$ plane:
  (a) Results calculated using the Ex-RDFT with PKO2 \cite{long2008_EEL82_12001};
  (b) same as (a), but with exchange energies excluded;
  (c) Results \cite{Yang2021_PRC104.054312,Yang2023_PRC107_024308,PCPK1_PES} obtained using the relativistic Hartree density functional PC-PK1 \cite{Zhao2010_PRC82.054319}.
  The energies are normalized with respect to the ground-state energies (indicated by red stars).
  The energy interval between the neighboring contour lines is 0.5 MeV.
  }
  \label{fig:Ru}
\end{figure*}

With the present Ex-RDFT framework, we can also study triaxial nuclei, which is currently beyond the capacity of the RHF method.  
The triaxial shapes of the Ru isotopes, in particular with $N \sim 66$ situated in the middle of the shell, are among the most intriguing nuclear phenomena. 
Nuclear shape phase transition, shape coexistence, and triaxial ground states are expected to be observed.
Both experimental and theoretical studies have shown the $\gamma$ softness around this region \cite{Stachel1982_NPA383_429, Stachel1984_ZPA316_105, Aysto1990_NPA515_365, Shannon1994_PLB336_136, Soderstrom2013_PRC88_024301, Abusara2017_PRC95_054302, Shi2018_PRC97_034329, Yang2023_PRC107_024308}.
In the following, we present the first study of triaxial nuclei within the framework of the relativistic energy density functional theory with exchange correlations. 
Here, the pairing forces are adjusted to reproduce the experimental pairing gap of $^{104}\text{Ru}$, estimated using the five-point formula \cite{bender2000_EPJA8_59a}.
The obtained potential energy surfaces (PESs) of the neutron-rich Ru isotopes against the deformation parameters $(\beta,\gamma)$ are depicted in Fig. \ref{fig:Ru} in comparison with the PKO2 results obtained by excluding the exchange energyies and the results \cite{Yang2021_PRC104.054312, Yang2023_PRC107_024308, PCPK1_PES} given by the relativistic Hartree density functional PC-PK1 \cite{Zhao2010_PRC82.054319}.

Both the PKO2 and PC-PK1 calculations present a rapid shape evolution from $^{104}\text{Ru}$ to $^{120}\text{Ru}$.
It can be clearly seen that the PESs are notably softened in the PKO2 results, which include the exchange correlations.
For $^{112}\text{Ru}$, in particular, the PKO2 PES is much softer in the $\gamma$ direction than the PC-PK1 PES.
To clarify the impact of the exchange correlations, we exclude the exchange energies from the PKO2 results and present the corresponding PESs in the panel (b) of Fig. 3.
It is evident that the exchange correlations significantly soften the PESs of these Ru isotopes.
It is worthwhile to mention that the $\pi$-exchange and $\rho$-tensor couplings also influence the potential energy curves in the previous axially deformed RHF calculations \cite{geng2020Phys.Rev.C101.064302, geng2022Phys.Rev.C105.034329, Geng2023_CPC47_044102}.
It will be interesting to study their contributions to nuclear triaxial softness within the Ex-RDFT in the future. 
Further studies employing beyond-mean-field methods are necessary to obtain the low-lying spectrum, enabling a direct comparison with the experimental data and clarifying the differences between the density functionals with and without exchange correlation energies.

In conclusion, the exact-exchange relativistic density functional theory has been solved for the first time in three-dimensional coordinate space using the inverse Hamiltonian and the Fourier spectral methods.
The orbital-dependent relativistic exchange energy density functional is constructed with the effective interaction PKO2, and the local exchange potentials are obtained by employing the ROEP method with the RKLI approximation.
The proposed framework has been benchmarked against the results obtained using the traditional RHF method for spherical doubly magic nuclei and axially deformed nuclei. 
The results demonstrate that the proposed framework is highly accurate.
The developed framework has been employed to study the triaxial deformation of neutron-rich Ru isotopes, which is beyond the capacity of the RHF method. 
The Ex-RDFT calculations with PKO2 predict the $\gamma$-softness of the Ru isotopes around $N\sim 44$, which is in good agreement with the experimental observations.
The present novel approach establishes a foundation for the study of nuclei without imposing any symmetry restrictions employing a relativistic density functional with exchange correlations.
The incorporation of $\pi$-exchange and $\rho$-tensor couplings in Ex-RDFT is currently underway with the objective of understanding the tensor-force effects.

\section*{Acknowledgments}
The authors thank J. Geng and W.H. Long for providing the RHF results of $^{20}$Ne, and Y.L. Yang for providing the Ru PESs calculated with PC-PK1.
This work is supported by the Institute for Basic Science (Grant No. IBS-R031-D1), 
the National Natural Science Foundation of China (Grants No. 12070131001, No. 11935003, No.11975031, and No. 12141501),
the European Research Council (ERC) under the European Union's Horizon 2020 research and innovation programme (Grant agreement No. 101018170),
JSPS KAKENHI (Grant No. JP23H05434), and JST ERATO (Grant No. JPMJER2304).
Computational works for this research are performed on the RCNP Computer System for Nuclear Physics and the High Performance Computing Resources in the IBS Research Solution Center.




\clearpage

\end{document}